\documentclass[]{spie}

\usepackage{amsmath,amsfonts,amssymb}
\usepackage{graphicx}
\usepackage[colorlinks=true, allcolors=blue]{hyperref}
\usepackage{svg}
\usepackage{subcaption}
\usepackage{siunitx}

\title{Design and modeling of an oblique incidence dichroic mirror for a high contrast imaging system}

\author[a]{Joanna Rosenbluth}
\author[b]{Ramya Anche}
\author[b]{Kyle Van Gorkom}
\author[b]{Nikoli Cooper}
\author[a]{William Melby}
\author[a,b]{Daewook Kim}
\author[b]{Ewan Douglas}

\affil[a]{Wyant College of Optical Sciences, University of Arizona, \linebreak 1630 E University Blvd, Tucson, AZ 85721, USA}
\affil[b]{Steward Observatory, University of Arizona, 933 N Cherry Ave, Tucson, AZ 85721, USA}

\authorinfo{Further author information: (Send correspondence to J.R.)}

\begin{document} 
\maketitle

\begin{abstract}

The proposed Habitable Worlds Observatory (HWO) aims to detect and characterize Earth-like planets around Sun-like stars from the ultraviolet to the infrared using high-contrast imaging. A dichroic mirror is a natural choice for splitting light between multiple wavelength channels in such a system, but the coating must introduce minimal polarization aberrations and chromatic wavefront errors (WFE) to avoid limiting contrast. We present the design of a long-pass dichroic mirror operated at a \(7^\circ\) angle of incidence that provides high reflection in the short-wavelength band, high transmission in the long-wavelength band, and a sharp cut-on near 550~nm. A multilayer thin-film design process was used to meet these spectral requirements while reducing polarization aberrations and chromatic WFE sensitivity. Reflected phase optimization was also incorporated to reduce sensitivity to manufacturing-induced layer-thickness variations, which can produce wavelength-dependent WFE and degrade coronagraph performance. The resulting coating-induced WFE was propagated through simulations of a vector vortex coronagraph to assess contrast performance. Simulations predict dark-hole contrast residuals at the \(10^{-12}\) level, indicating that the optimized dichroic mirror does not limit coronagraph performance.

\end{abstract}


\section{INTRODUCTION}

Future space-based observatories such as the Habitable Worlds Observatory (HWO) aim to directly image and characterize Earth-like exoplanets around nearby Sun-like stars using high-contrast imaging.\cite{feinberg_habworlds}  Achieving the required contrast levels places stringent requirements on the optical system, including tight control of wavefront quality, stray light, and polarization-dependent aberrations. Even small amplitude or phase differences between orthogonal polarization states can introduce leakage into the coronagraph dark hole and degrade achievable contrast.\cite{sayson_vvc,willthesis,Breckinridge_2015,Anche_2023} As coronagraph architectures continue to push toward deeper contrast and broader spectral coverage, controlling polarization effects throughout the optical system becomes increasingly important.

Many proposed coronagraph instruments operate over broad wavelength ranges spanning much of the visible spectrum. However, wavefront sensing and coronagraph suppression become more challenging over large spectral bandwidths because chromatic phase errors and polarization aberrations vary with wavelength.\cite{mawet_wvl_contrast} Splitting the beam into multiple narrower spectral channels can reduce these chromatic effects while still providing broad overall wavelength coverage.\cite{hwo_dichroicspecs,pueyo2019} Dichroic beamsplitters are well suited for this task because they separate wavelength bands with minimal throughput loss, directing different portions of the spectrum into independent optical channels rather than rejecting light.\cite{Stark_2024} For this reason, dichroic coatings are a promising solution for wavelength separation in future high-contrast imaging instruments.\cite{zenodoslides}

Although dichroic coatings provide efficient wavelength separation, multilayer coatings operated at oblique incidence inherently exhibit polarization-dependent behavior because s- and p-polarized light experience different Fresnel boundary conditions. These differences accumulate across the many interfaces within a multilayer coating and produce measurable diattenuation and retardance.\cite{harrington_dkist2} For example, Mueller-matrix measurements of a commercial dichroic operated at 45$^\circ$ incidence showed strong wavelength-dependent polarization effects, with the optic behaving approximately as a quarter-wave plate near 630~nm.\cite{meredith_dichroic} Previous studies of the dichroic beamsplitter used to separate the VIS and NISP channels in the Euclid space telescope showed that coating-induced phase variations can generate strongly chromatic wavefront errors capable of distorting the telescope point-spread function.\cite{baron2024dichroic} Measurements of this optic reported wavelength-dependent wavefront distortions approaching 40 nm RMS at specific wavelengths, highlighting the importance of accurately controlling and characterizing coating-induced phase effects.\cite{euclid_measure} These effects become increasingly significant as the angle of incidence  (AOI) increases, motivating the use of a small AOI for dichroic designs used in applications requiring precise polarization control.\cite{habexprt}

Beyond conventional throughput requirements, coating-induced phase errors can directly impact coronagraph performance. Spatial and spectral variations in the coating phase response generate chromatic wavefront errors that are difficult to correct over broad bandwidths and can produce residual light leakage within the coronagraph dark hole. Understanding the relationship between coating design, wavefront error, and achievable contrast is therefore an important consideration for future high-contrast imaging systems.

In this work, a 7° AOI dichroic mirror is designed for use in a high-contrast imaging instrument. The resulting coating is analyzed in terms of its spectral performance, polarization aberrations, and chromatic wavefront error. Several phase-based optimization strategies are investigated to reduce sensitivity to coating thickness variations. The resulting coating-induced chromatic wavefront error is propagated through vector vortex coronagraph simulations to assess contrast performance. Comparisons are made between a baseline design optimized using conventional spectral criteria and a design incorporating reflected phase optimization to quantify the impact of phase control on coronagraph contrast. Sec.~\ref{sec:background2} reviews the thin-film, polarization, and chromatic wavefront-error background relevant to high-contrast coronagraph systems. Sec.~\ref{sec:design3} describes the dichroic coating design methodology and phase optimization strategies. Sec.~\ref{sec:perf4} presents the predicted coating performance, and Sec.~\ref{sec:contrast5} evaluates the resulting impact on coronagraph contrast through vector vortex coronagraph simulations.


\section{BACKGROUND}
\label{sec:background2}

Multilayer thin-film coatings rely on interference between partial reflections at each interface to achieve the desired spectral performance.\cite{macleod} At oblique incidence, this interference becomes polarization-dependent due to the different boundary conditions experienced by s- and p-polarized light.\cite{chipman} As a result, both the amplitude and phase behavior of the coating must be evaluated separately for each polarization state.

Two coating-induced effects are of particular importance for high-contrast coronagraph systems. The first is polarization aberration, which arises from differences between the s- and p-polarization states and manifests as diattenuation and retardance. The second is chromatic wavefront error (WFE), which results from nonuniformities of the coating and the wavelength dependence of the coating phase response. Because these errors vary with wavelength, they can leave chromatic residuals after broadband wavefront correction and degrade dark-hole contrast.\cite{kyle_uv_jatis,anche2023} Therefore, polarization aberrations and chromatic WFE must be considered alongside conventional spectral performance metrics when designing dichroic coatings for high-contrast coronagraph systems.

\subsection{Thin-film modeling using the characteristic matrix method}

The optical response of the multilayer coating was modeled using the characteristic matrix method.\cite{macleod} For each layer, the phase thickness is given by

\begin{equation}
\delta_j = \frac{2\pi}{\lambda} n_j d_j \cos\theta_j,
\end{equation}

where $n_j$, $d_j$, and $\theta_j$ are the refractive index, physical thickness, and propagation angle within the layer, respectively.

The optical admittance is a polarization-dependent quantity which is given by

\begin{equation}
\eta_j^{(s)} = n_j \cos\theta_j,
\qquad
\eta_j^{(p)} = \frac{n_j}{\cos\theta_j}.
\end{equation}

These differing admittances are the source of the polarization-dependent behavior observed in multilayer coatings at oblique incidence.

Each layer is represented by a characteristic matrix

\begin{equation}
M_j =
\begin{bmatrix}
\cos\delta_j & \frac{i}{\eta_j}\sin\delta_j \\
i\eta_j\sin\delta_j & \cos\delta_j
\end{bmatrix}
\end{equation}

The overall response of the multilayer stack is obtained by multiplying the characteristic matrices of all layers and applying the substrate boundary condition

\begin{equation}
\begin{bmatrix}
B \\
C
\end{bmatrix}
=
\prod_j M_j
\begin{bmatrix}
1 \\
\eta_{\mathrm{sub}}
\end{bmatrix}.
\end{equation}

The effective optical admittance of the stack is then

\begin{equation}
Y=\frac{C}{B}.
\end{equation}

The resulting complex reflection and transmission coefficients are

\begin{equation}
r = \frac{\eta_0 - Y}{\eta_0 + Y},
\qquad
t = \frac{2\eta_0}{\eta_0 + Y},
\end{equation}

where $\eta_0$ is the optical admittance of the incident medium. From these coefficients, the reflectance, transmittance, and phase response of the coating can be calculated separately for s- and p-polarized light. The reflectance and transmittance are obtained from the complex amplitude coefficients as

\begin{equation}
R = |r|^2,
\qquad
T = \frac{\mathrm{Re}(\eta_{\mathrm{sub}})}
{\mathrm{Re}(\eta_0)}
|t|^2,
\end{equation}

where the admittance ratio accounts for the change in optical power between the incident and transmitted media. The phase response is determined from the argument of the complex coefficients,

\begin{equation}
\phi_r = \arg(r),
\qquad
\phi_t = \arg(t).
\end{equation}

\subsection{Polarization aberrations}

The polarization behavior of the coating is characterized using diattenuation and retardance.\cite{chipman} Diattenuation quantifies the differential attenuation of orthogonal polarization states and is evaluated separately for the reflected and transmitted beams as

\begin{equation}
D_{r} = \frac{R_{s} - R_{p}}{R_{s} + R_{p}},
\qquad
D_{t} = \frac{T_{s} - T_{p}}{T_{s} + T_{p}},
\end{equation}

where $R_s$, $R_p$, $T_s$, and $T_p$ are the reflectance and transmittance for s- and p-polarized light. A value of $D=0$ corresponds to identical response for both polarization states.

Retardance describes the relative phase difference between the two polarization components and is given by

\begin{equation}
\delta = \phi_p - \phi_s,
\end{equation}

where $\phi_s$ and $\phi_p$ are the phases of the complex amplitude coefficients for s- and p-polarization.

At normal incidence, s- and p-polarization are degenerate, resulting in zero diattenuation and retardance. As the angle of incidence increases, the Fresnel coefficients diverge, producing increasing amplitude and phase differences that accumulate throughout the multilayer stack and vary spectrally.\cite{harrington_dkist2}

\subsection{Coating phase response and wavefront error}

In addition to meeting spectral transmission and reflection requirements, the wavelength dependence of the coating phase response must also be considered. Multilayer interference coatings can exhibit regions where the reflected phase varies rapidly with wavelength, particularly near spectral transitions and resonant features.\cite{baumeister_book} In these regions, small coating thickness variations can produce large changes in reflected phase, making the coating highly sensitive to manufacturing nonuniformities.\cite{Ji_monitor}

For coronagraph applications, the quantity of interest is not the coating phase itself but the wavefront error generated from errors in layer thickness and spatial variation across the optic. These errors lead to wavelength-dependent phase errors that manifest as chromatic wavefront error (WFE).\cite{euclid_measure,euclid_prelimresults} A coating with a smooth phase response is generally less sensitive to thickness variations, whereas a coating with steep phase gradients can generate significant WFE even when spectral performance remains largely unchanged. This analysis focuses on the reflected phase because it exhibits significantly stronger spectral phase variation than the transmitted phase.\cite{piegari_wfe}

For a coating with spatially varying thickness errors, the reflected phase error can be expressed as

\begin{equation}
\Delta\phi_R(x,y,\lambda)
=
\phi_{R,\mathrm{actual}}(x,y,\lambda)
-
\phi_{R,\mathrm{nominal}}(\lambda),
\label{equ:dphi}
\end{equation}

where $\phi_{R,\mathrm{nominal}}$ is the phase response of the nominal coating design and $\phi_{R,\mathrm{actual}}$ is the phase response of the fabricated coating.\cite{baron2024dichroic} The corresponding coating-induced wavefront error is

\begin{equation}
\mathrm{WFE}_c(x,y,\lambda)
=
\frac{\lambda}{2\pi}
\Delta\phi_R(x,y,\lambda).
\label{equ:wfe}
\end{equation}

Consequently, the sensitivity of the coating to manufacturing nonuniformities is closely linked to the spectral behavior of the reflected phase response. Coatings with smoother phase characteristics are expected to exhibit reduced chromatic wavefront error when thickness variations are present. Because high-contrast coronagraph systems are highly sensitive to chromatic wavefront errors, the reflected phase response was considered as an additional design parameter during the optimization process.
\label{sec:phase_wfe}

\section{Dichroic Coating Design and Optimization}
\label{sec:design3}

\begin{figure} [ht]
   \begin{center}
   \begin{tabular}{c} 
   \includegraphics[height=6.5cm]{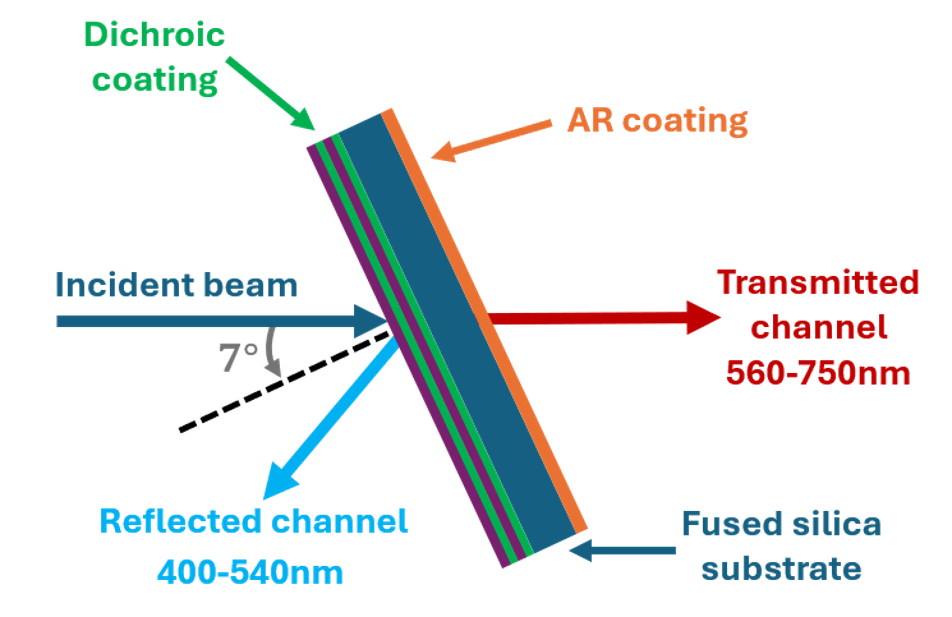}
   \end{tabular}
   \end{center}
   \caption[example] 
   { \label{fig:dichroic_concept} 
Architecture of a low-angle dichroic beamsplitter. A multilayer dielectric coating deposited on the front surface of a fused silica substrate reflects the 400--540~nm band while transmitting the 560--750~nm band. An antireflection coating on the rear surface minimizes backside reflections and maximizes throughput.}
   \end{figure}

The dichroic mirror was designed to satisfy the spectral throughput and polarization requirements of a multi-channel coronagraph instrument while maintaining high efficiency across the visible wavelength range.\cite{roy2026lazuli} In addition to meeting conventional reflection and transmission requirements, the design process considered polarization dependent amplitude and phase effects introduced by the multilayer coating. Phase-based optimization constraints
were also explored to reduce sensitivity to variations in coating thickness. Fig.~\ref{fig:dichroic_concept} illustrates the structure of the dichroic beamsplitter designed in this work.


\subsection{Design Requirements}

The dichroic beamsplitter must efficiently separate the visible spectrum into reflected and transmitted wavelength channels while preserving the high optical throughput required for coronagraphic observations. The primary spectral and manufacturing requirements used during coating optimization are summarized in Table~\ref{tab:dichroic_requirements}.

\begin{table}[ht]
\caption{Dichroic coating design requirements.} 
\label{tab:dichroic_requirements}
\begin{center}       
\begin{tabular}{|l|l|l|} 
\hline
\rule[-1ex]{0pt}{3.5ex} \textbf{Specification} & \textbf{Value} & \textbf{Notes}  \\
\hline
\rule[-1ex]{0pt}{3.5ex} Reflection band & 400--540~nm &  \\
\hline
\rule[-1ex]{0pt}{3.5ex} Transmission band & 560--750~nm &  \\
\hline
\rule[-1ex]{0pt}{3.5ex} Transition region & 540--560~nm & Cut-on near 550~nm \\
\hline
\rule[-1ex]{0pt}{3.5ex} Reflectance & $R_s, R_p > 98$\%& As-built: $R_s, R_p > 95$\% \\
\hline
\rule[-1ex]{0pt}{3.5ex} Transmittance & $T_s, T_p > 98$\%& As-built: $T_s, T_p > 95$\% \\
\hline
\rule[-1ex]{0pt}{3.5ex} Angle of incidence & 7$^\circ$ & Collimated space \\
\hline
\rule[-1ex]{0pt}{3.5ex} Substrate & Fused silica & 6~mm thickness \\
\hline
\rule[-1ex]{0pt}{3.5ex} Backside AR coating & $<$2\% reflectance & 560--750~nm \\
\hline
\rule[-1ex]{0pt}{3.5ex} Diameter & 25.4~mm & 90\% clear aperture \\
\hline
\end{tabular}
\end{center}
\end{table}

A fused silica substrate and nominal incidence angle of 7$^\circ$ were selected for the design. The selected angle provides sufficient separation between the incident and reflected beam paths for instrument implementation while maintaining low polarization sensitivity due to the small angle of incidence. Additional design considerations included minimizing backside reflections through the use of an anti-reflection coating optimized for the transmission band.

\subsection{Coating Architecture and Material Selection}

The dichroic design starting point was based on the periodic multilayer stack

\begin{equation}
(0.5H\,L\,0.5H)^n,
\end{equation}

where $H$ and $L$ represent quarter-wave optical thickness layers of the high- and low-index materials, respectively, and $n$ is the number of repeated periods. This structure is a standard edge-filter design and serves as a convenient foundation for optimization of a dichroic mirror.

\begin{figure}[ht]
  \centering
  \begin{subfigure}[t]{0.5\textwidth}
    \centering
    \includegraphics[width=1\textwidth]{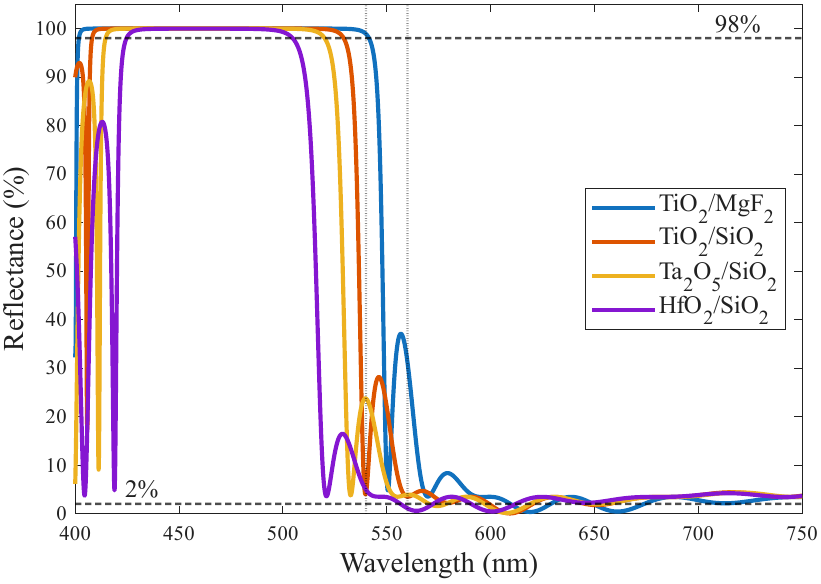}
    \subcaption{ }
    \label{fig:material_comparison_a}
  \end{subfigure}\hfill
  \begin{subfigure}[t]{0.5\textwidth}
    \centering
    \includegraphics[width=1\textwidth]{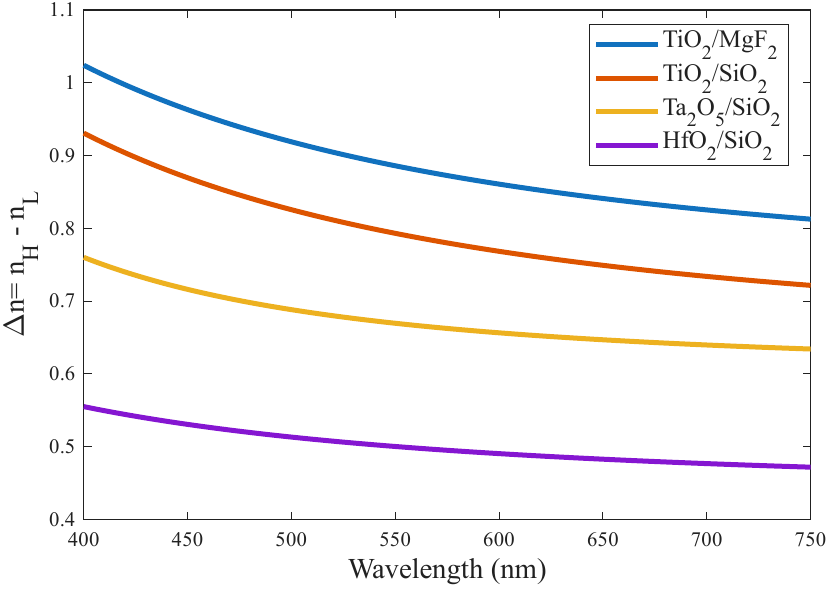}
    \subcaption{ }
    \label{fig:material_comparison_b}
  \end{subfigure}\hfill
  \caption{(a) Nominal spectral performance of several candidate material pairs using the same periodic multilayer architecture prior to optimization.
(b) Refractive-index difference, $n_H-n_L$, for each material pair as a function of wavelength.}
  \label{fig:material_comparison}
\end{figure}

Several candidate material combinations were evaluated using this same initial multilayer architecture. Fig.~\ref{fig:material_comparison_a} compares the spectral performance for each pair of materials prior to numerical optimization, while Fig.~\ref{fig:material_comparison_b} shows the corresponding refractive-index difference between the high and low-index materials. As expected, material pairs with larger refractive-index contrast produce broader reflection bands and steeper spectral transitions. The TiO$_2$/MgF$_2$ design achieves high reflectivity across the entire 400–540 nm reflection band. The lower-contrast material pairs exhibit progressively narrower stopbands. This trend is particularly evident for the HfO$_2$/SiO$_2$ design, which only achieves $>98\%$ reflectivity over a bandwidth of 80~nm, far less than the 140~nm requirement. These results highlight the importance of high refractive-index contrast for broadband dichroic mirrors.

The TiO$_2$/MgF$_2$ material pair provides the largest refractive-index contrast among the combinations considered and correspondingly exhibits the strongest initial spectral performance. However, SiO$_2$ was selected as the low-index material for the final design because it offers improved manufacturability and robustness to environmental factors. Previous studies have noted that MgF$_2$ films can be more susceptible to mechanical stress and environmental degradation, while SiO$_2$ offers improved chemical stability and resistance to stress, humidity, and temperature cycling.\cite{ghapanvari_sio2_mgf2_2025} These considerations are especially important when a coating will be used in space. Based on this trade-off between optical performance and manufacturability, the final optimization used TiO$_2$ as the high-index material and SiO$_2$ as the low-index material. The resulting design was optimized in \textit{OptiLayer} to meet the spectral requirements while reducing the polarization-dependent amplitude and phase effects.

\subsection{Phase Optimization for High-Contrast Imaging}

Although the spectral requirements define the primary functionality of the dichroic mirror, the reflected spectral phase response was also considered during the optimization process, as mentioned in Sec.~\ref{sec:phase_wfe}. Initial designs that satisfied the reflection and transmission requirements often exhibited sharp phase features within the reflection band. These features correspond to regions where the reflected phase changes rapidly with wavelength and are expected to be more sensitive to coating thickness variations. This would induce a larger and more spectrally varying WFE, as shown in Equations \ref{equ:dphi} and \ref{equ:wfe}. Because chromatic wavefront errors are difficult to correct in broadband coronagraph systems, additional optimization constraints were investigated to improve the smoothness of the reflected phase response. \textit{OptiLayer}'s gradual evolution optimization method was used, with equal weighting between the spectral and phase targets.\cite{opt_method}

\begin{figure}[ht]
\begin{center}
\includegraphics[width=0.55\textwidth]{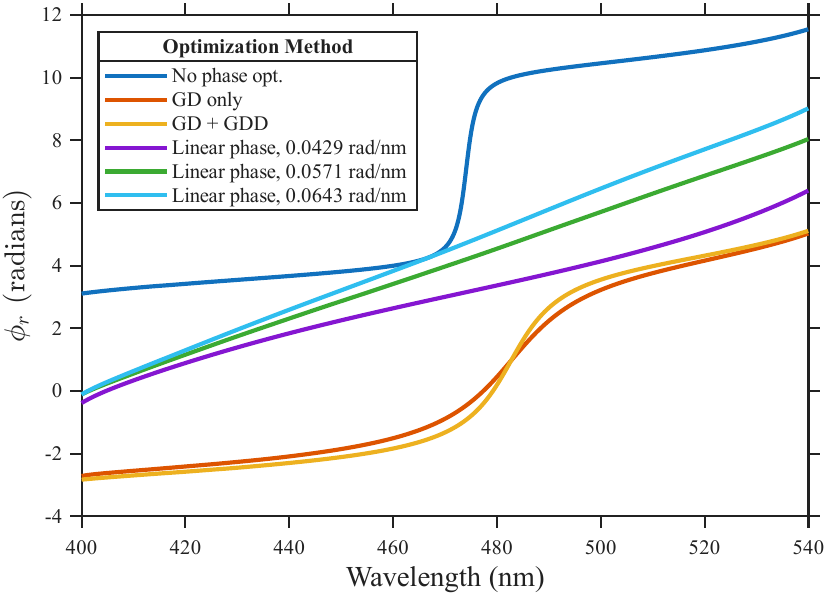}
\end{center}
\caption{
Reflected phase response across the reflection band for the investigated phase optimization strategies. The design optimized using only reflection and transmission targets exhibits a sharp phase modulation near the center of the reflection band, while the linear phase targets produce a substantially smoother phase response.
}
\label{fig:phase_optimization}
\end{figure}

\begin{figure}[ht]
  \centering
  \begin{subfigure}[t]{0.5\textwidth}
    \centering
    \includegraphics[width=1\textwidth]{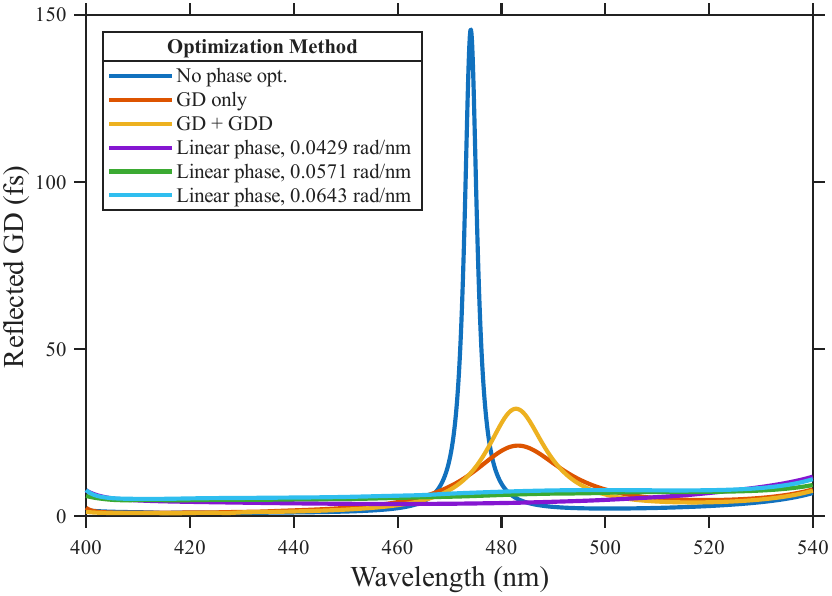}
    \subcaption{}
    \label{fig:gd_compare}
  \end{subfigure}\hfill
  \begin{subfigure}[t]{0.5\textwidth}
    \centering
    \includegraphics[width=1\textwidth]{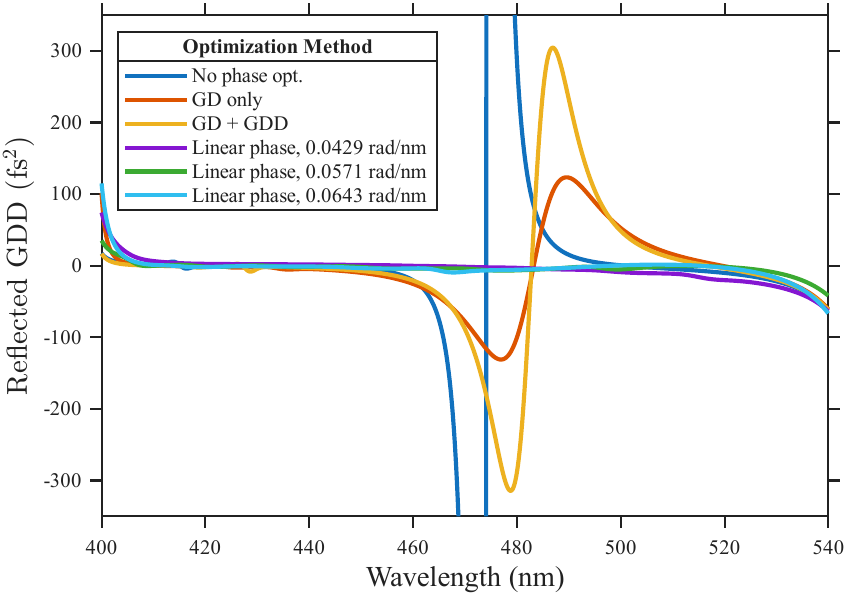}
    \subcaption{}
    \label{fig:gdd_compare}
  \end{subfigure}
  \caption{
Comparison of reflected (a) group delay (GD) and (b) group delay dispersion (GDD) for the investigated phase optimization strategies. Optimization toward a linear reflected phase target produces a more uniform GD and GDD response than directly optimizing GD or GD+GDD simultaneously.
}
  \label{fig:gd_gdd_comparison}
\end{figure}

Several phase-based optimization strategies were evaluated. These included constraining the reflected group delay (GD) and group delay dispersion (GDD), which correspond to the first and second spectral derivatives of the reflected phase, respectively, as well as optimization toward an explicit linear reflected phase response. Multiple linear phase targets with different slopes were investigated to evaluate the tradeoff between phase linearity and manufacturing sensitivity.

\begin{figure}[ht]
  \centering

  \begin{subfigure}[t]{0.85\textwidth}
    \centering
    \includegraphics[width=\textwidth]{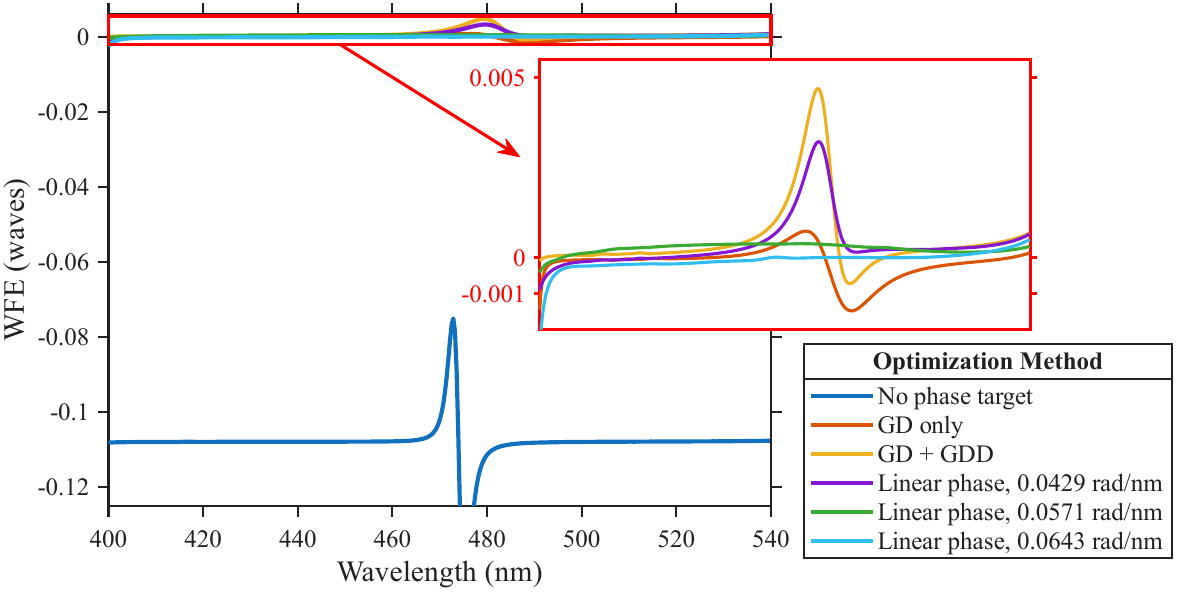}
    \subcaption{}
    \label{fig:wfe_compare}
  \end{subfigure}

  \begin{subfigure}[t]{0.85\textwidth}
    \centering
    \includegraphics[width=\textwidth]{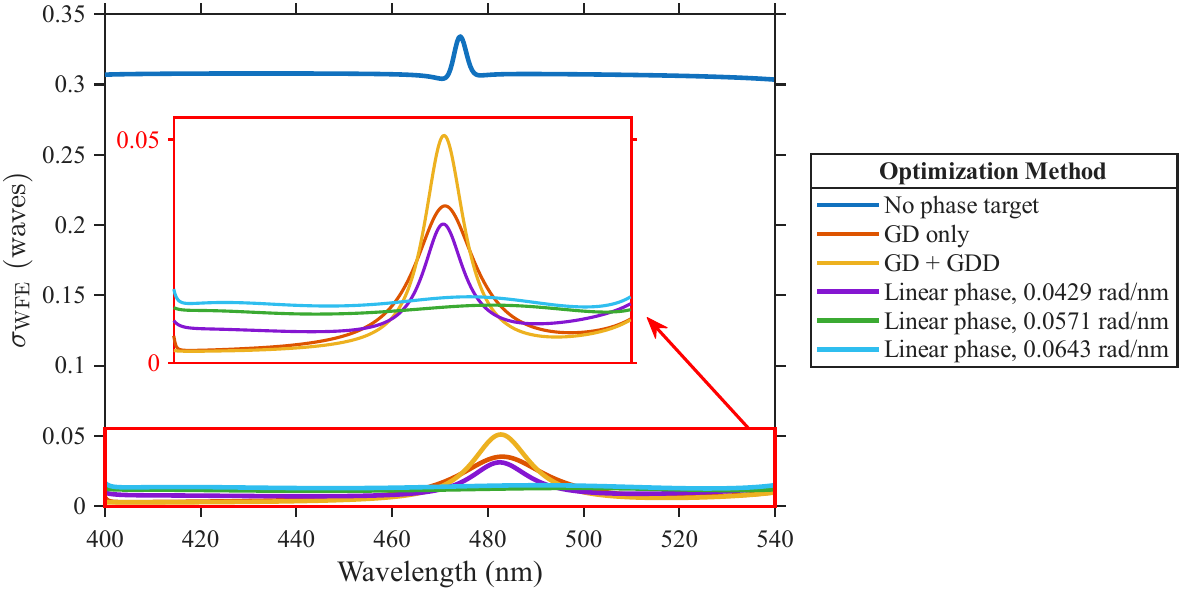}
    \subcaption{}
    \label{fig:sig_compare}
  \end{subfigure}

  \caption{
Comparison of (a) WFE and (b) WFE standard deviation between different optimization methods. Without optimizing for reflected phase, these values are large in magnitude and vary rapidly with wavelength. 
}
  \label{fig:wfe_comparison}
\end{figure}

Fig.~\ref{fig:phase_optimization} compares the reflected phase response obtained using the different optimization strategies. The design optimized using only reflection and transmission targets exhibits a sharp phase modulation near the center of the reflected band. Although GD- and GD+GDD-based merit functions reduce the severity of this feature, the most uniform phase response is obtained using a linear phase target. The corresponding GD and GDD are shown in Fig.~\ref{fig:gd_gdd_comparison}. Optimization toward a linear reflected phase target produces the smoothest GD and GDD behavior, indicating that directly controlling the phase is more effective than independently constraining its derivatives.

The resulting phase behavior was further evaluated through Monte Carlo analysis to estimate sensitivity to coating thickness perturbations. Further details of the tolerancing procedure are provided in Sec.~\ref{sec:montecarlo}. WFE and $\sigma_{\mathrm{WFE}}$ were calculated from the difference between the expected and nominal reflected phase responses and the associated 68\% confidence interval. The results are shown in Fig.~\ref{fig:wfe_comparison}. The design optimized without a reflected phase target exhibits WFE and $\sigma_{\mathrm{WFE}}$ values that are more than an order of magnitude larger than the phase-optimized designs. Optimization using GD and GD+GDD significantly reduces these metrics, but both approaches retain pronounced spectral features near 485~nm. In contrast, the linear phase targets substantially suppress these spectral variations, demonstrating that phase linearity is closely linked to reduced manufacturing sensitivity. Among the linear phase designs, larger target slopes produced lower spectral variation but increased the overall magnitude of WFE and $\sigma_{\mathrm{WFE}}$. A target slope of 0.0571~rad/nm provided the best balance between minimizing chromatic variation and maintaining low overall wavefront error and was therefore selected for the final design.

\label{sec:phaseopt}

\newpage

\section{Predicted Coating Performance}
\label{sec:perf4}

The final coating design consists of a 47-layer TiO$_2$/SiO$_2$ dichroic coating deposited on a 6~mm thick fused silica substrate with a 5-layer anti-reflection coating on the backside. In addition to optimizing for the linear phase, manufacturing constraints such as having a minimum layer thickness of 20 nm were considered. Following optimization, the coating performance was evaluated using spectral, polarization, and phase-based metrics relevant to high-contrast coronagraph applications. 

\subsection{Spectral Performance}

\begin{figure}[ht]
\begin{center}
\includegraphics[width=0.6\textwidth]{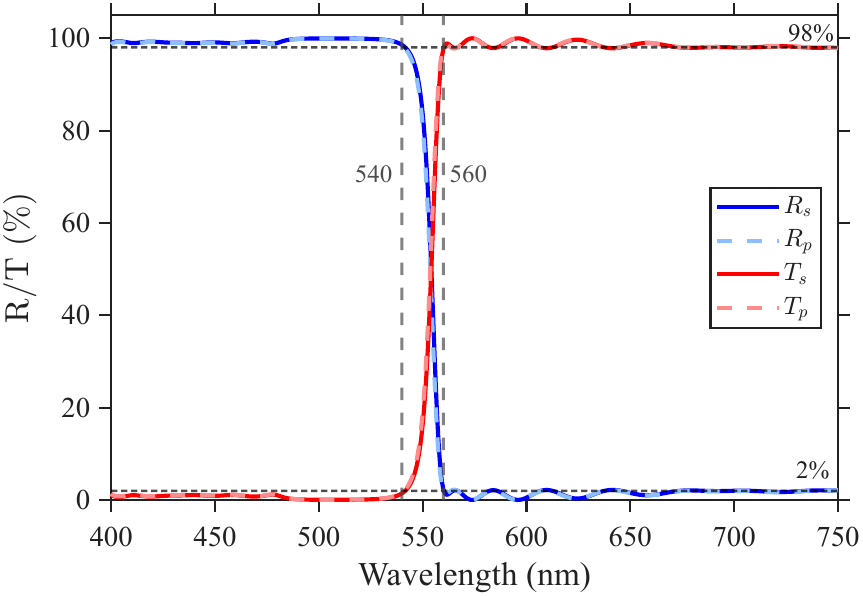}
\end{center}
\caption{
Spectral performance of the final optimized dichroic coating at $7^\circ$ angle of incidence. 
}
\label{fig:final_spectral_performance}
\end{figure}

Fig.~\ref{fig:final_spectral_performance} shows the predicted spectral performance of the final optimized design. Reflectance and transmittance remain above 95\% throughout the specified operating bands, with peak values approaching 98\%. This comfortably satisfies the performance requirements defined in Table \ref{tab:dichroic_requirements}. While quarter-wave antireflection coatings are frequently used, our goals of minimizing diattenuation and maximizing throughput in the transmitted band require a more robust 5-layer design. This design uses the same high- and low-index materials as the front dichroic coating to simplify the manufacturing process and achieves back reflectance \(<2\%\) throughout the transmitted band.

\subsection{Polarization Performance}

\begin{figure}[ht]
  \centering
  \begin{subfigure}[t]{0.49\textwidth}
    \centering
    \includegraphics[width=\textwidth]{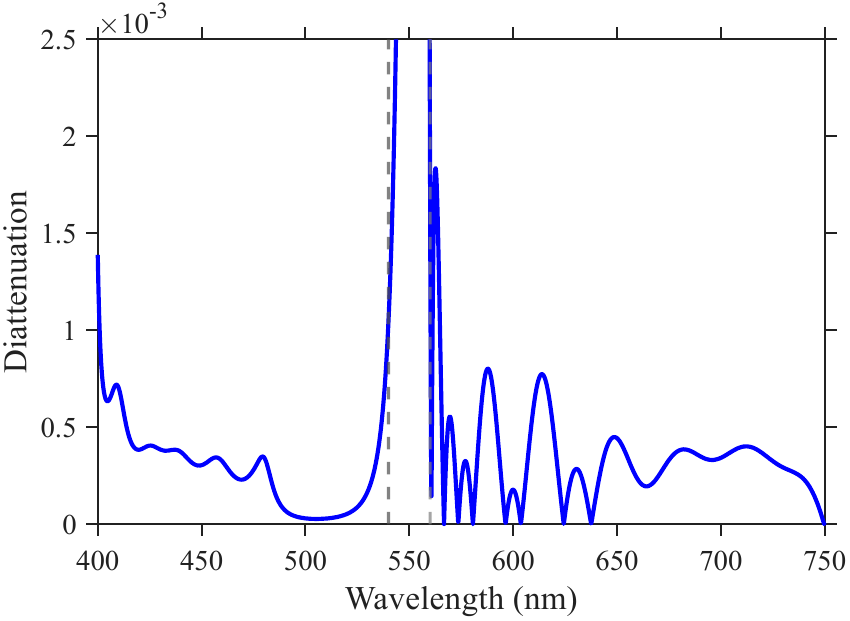}
    \subcaption{}
    \label{fig:final_pol_a}
  \end{subfigure}\hfill
  \begin{subfigure}[t]{0.49\textwidth}
    \centering
    \includegraphics[width=\textwidth]{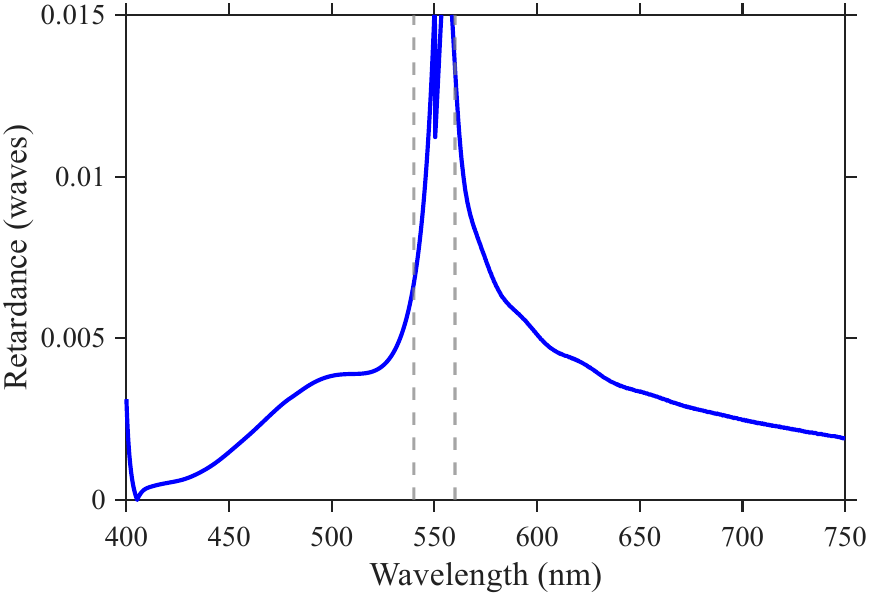}
    \subcaption{}
    \label{fig:final_pol_b}
  \end{subfigure}

  \caption{
Polarization properties of the final dichroic design. (a) Diattenuation fluctuates between 0 and 0.001 throughout the operating wavelength region. (b) Analysis predicts retardance values \(<0.008\) waves and \(<0.015\) waves in the reflection and transmission regions, respectively. Dashed lines indicate the transition region. 
}
\label{fig:final_polarization}
\end{figure}

The polarization performance of the final design is summarized in Fig.~\ref{fig:final_polarization}. As expected for a low-AOI dichroic mirror, both quantities remain relatively small across most of the operating wavelength range with the largest deviations occurring near the spectral transition region. Optimizing for s-pol and p-pol individually proves to be an effective way of further minimizing polarization effects. Therefore, coating-induced polarization aberrations are not expected to limit contrast in our system.

\newpage
\subsection{Reflected Phase Characteristics}

\begin{figure}[ht]
  \centering
  \begin{subfigure}[t]{0.49\textwidth}
    \centering
    \includegraphics[width=\textwidth]{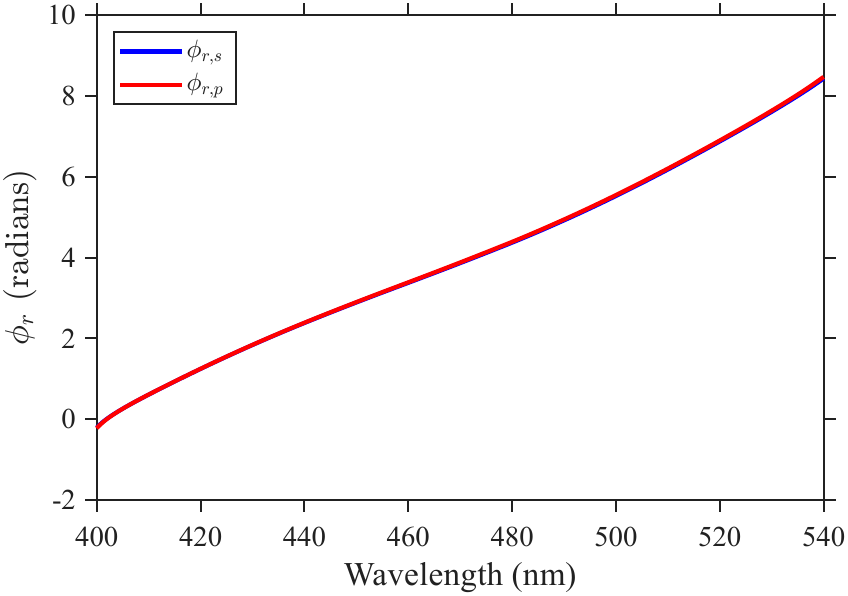}
    \subcaption{}
    \label{fig:final_phase}
  \end{subfigure}\hfill
  \begin{subfigure}[t]{0.49\textwidth}
    \centering
    \includegraphics[width=\textwidth]{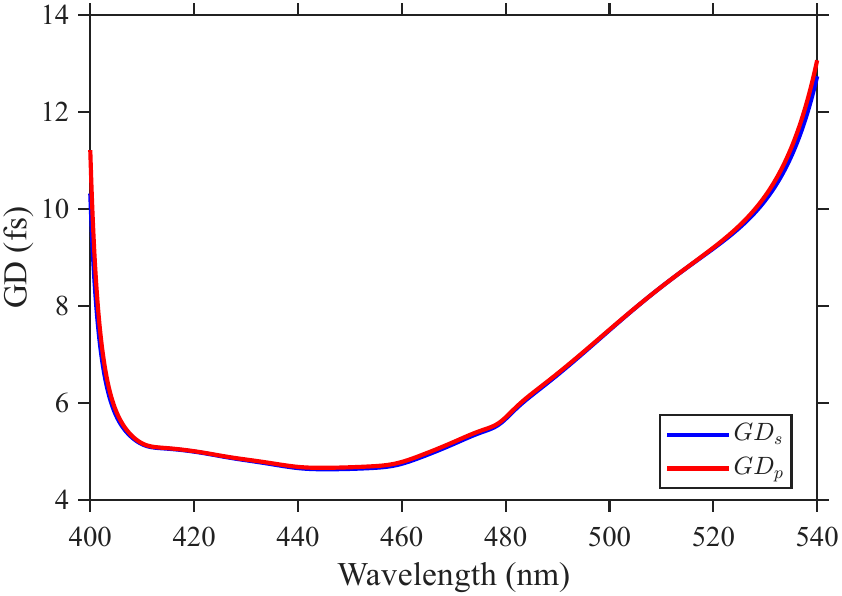}
    \subcaption{}
    \label{fig:final_gd}
  \end{subfigure}

  \caption{
The reflected phase (a) follows a linear relationship with wavelength for both s- and p-polarizations. GD values (b) remain under 14~fs and vary 8~fs throughout the reflection band. 
}
\label{fig:final_phase_gd}
\end{figure}

The reflected phase response of the final coating is shown in Fig.~\ref{fig:final_phase_gd}. The phase varies smoothly across the reflection band and closely follows the linear behavior targeted during optimization. The corresponding group delay is also smooth and exhibits only small differences between s- and p-polarized light. These results indicate that the phase optimization strategy successfully suppressed the sharp spectral phase features observed in the initial designs while maintaining minimal polarization-dependent phase behavior.

\subsection{Manufacturing Sensitivity and Wavefront Error}

\begin{figure}[ht]
  \centering

  \begin{subfigure}[t]{0.5\textwidth}
    \centering
    \includegraphics[width=1\textwidth]{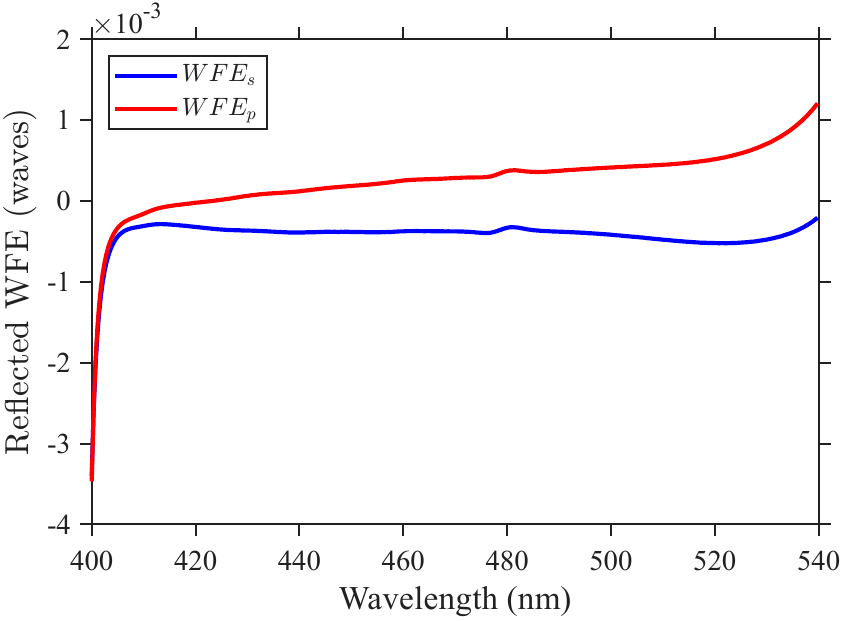}
    \subcaption{}
    \label{fig:final_wfe_a}
  \end{subfigure}\hfill
  \begin{subfigure}[t]{0.5\textwidth}
    \centering
    \includegraphics[width=1\textwidth]{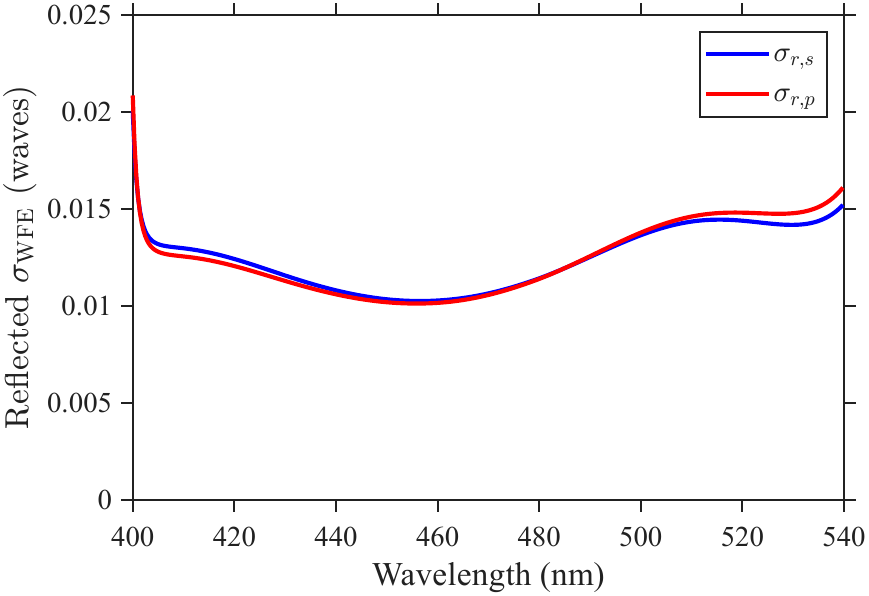}
    \subcaption{}
    \label{fig:final_wfe_b}
  \end{subfigure}

    \begin{subfigure}[t]{0.5\textwidth}
    \centering
    \includegraphics[width=1\textwidth]{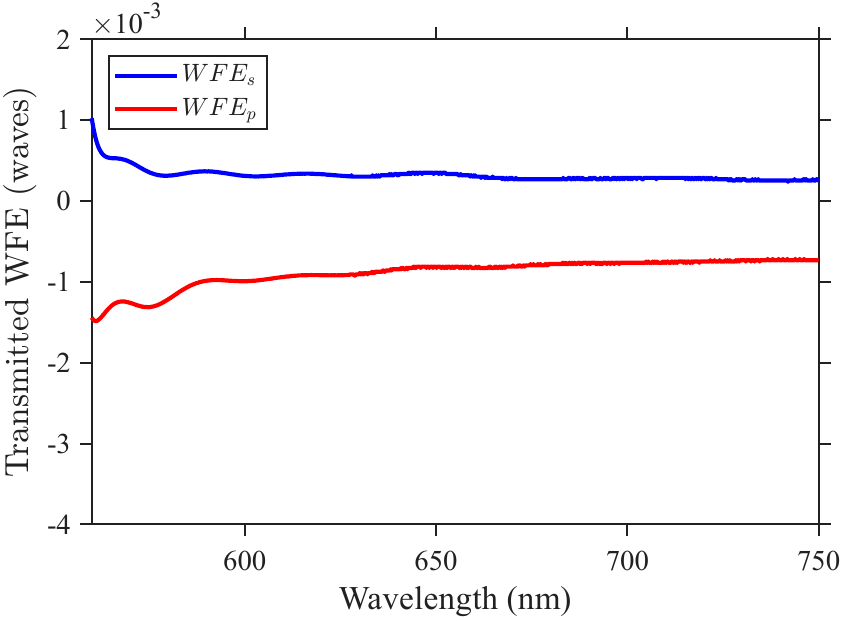}
    \subcaption{}
    \label{fig:final_wfe_c}
  \end{subfigure}\hfill
  \begin{subfigure}[t]{0.5\textwidth}
    \centering
    \includegraphics[width=1\textwidth]{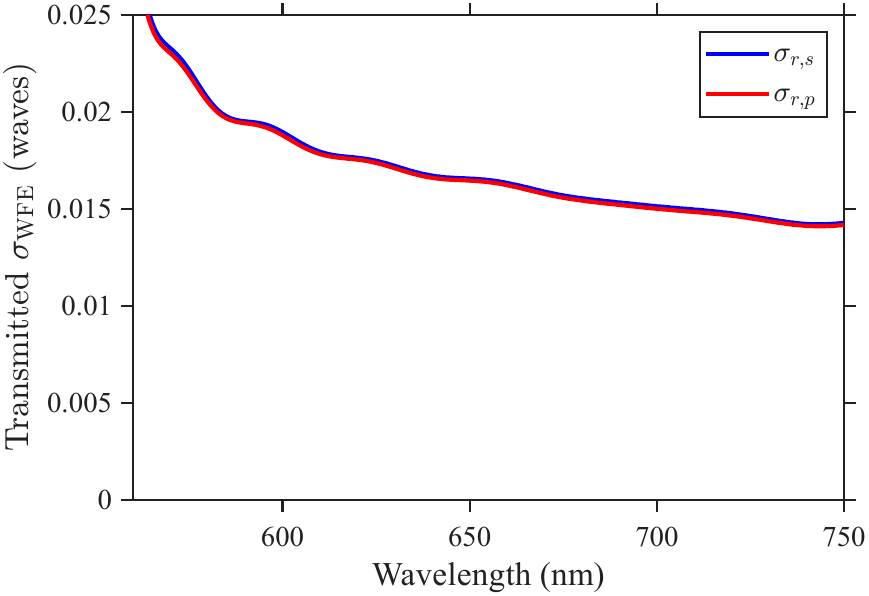}
    \subcaption{}
    \label{fig:final_wfe_d}
  \end{subfigure}

  \caption{
Wavefront-error metrics derived from the Monte Carlo analysis of the final coating design using 1,000 trials with 1\% layer-thickness errors. (a) Coating-induced wavefront error relative to the nominal design. (b) One-sigma wavefront-error uncertainty as a function of wavelength. Corresponding metrics for the transmitted band are shown in (c) and (d). The wavefront-error variation remains smooth across the reflection band and exhibits no sharp resonant features.
}
\label{fig:final_wfe}
\end{figure}

To evaluate the sensitivity of the final coating to fabrication errors, a Monte Carlo analysis consisting of 1,000 trials was performed using independent 1\% layer-thickness perturbations. Fig.~\ref{fig:final_wfe} shows the resulting coating-induced WFE and one-sigma WFE uncertainty derived from the reflected phase response for the reflected and transmitted beams.

The mean coating-induced WFE magnitude remains below 0.003 waves across the reflection and transmission bands for both polarization states. More importantly, the associated WFE uncertainty remains smooth and free of the strong spectral features observed in the earlier optimization studies. The absence of localized peaks indicates that the reflected phase response is relatively insensitive to small thickness perturbations and suggests that the final design is more robust to manufacturing variability. Broadband optical monitoring techniques may also provide a practical means of characterizing and mitigating deposition errors during fabrication.\cite{Tikhonravov_monitoring,Ji_monitor}

\begin{figure}[ht]
  \centering

  \begin{subfigure}[t]{0.5\textwidth}
    \centering
    \includegraphics[width=\textwidth]{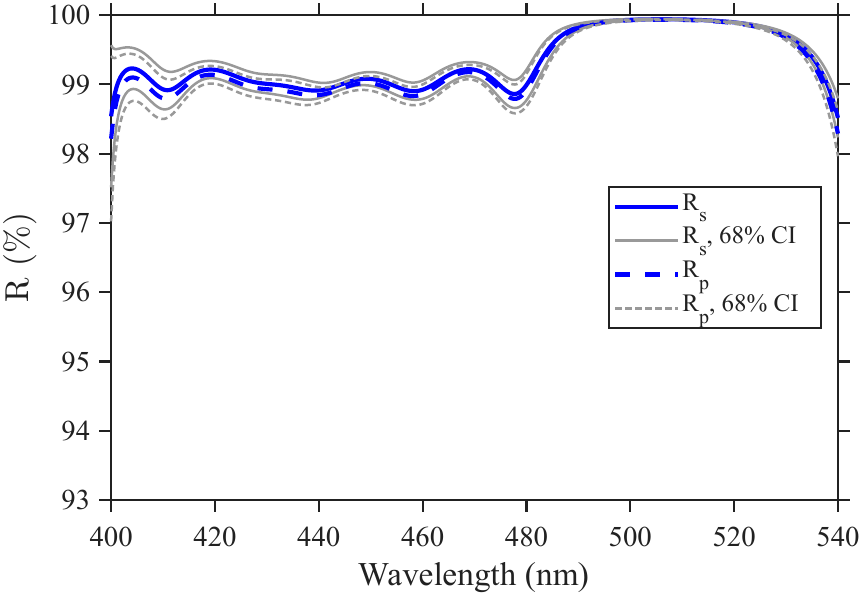}
    \subcaption{}
    \label{fig:final_montecarlo_r}
  \end{subfigure}\hfill
  \begin{subfigure}[t]{0.5\textwidth}
    \centering
    \includegraphics[width=\textwidth]{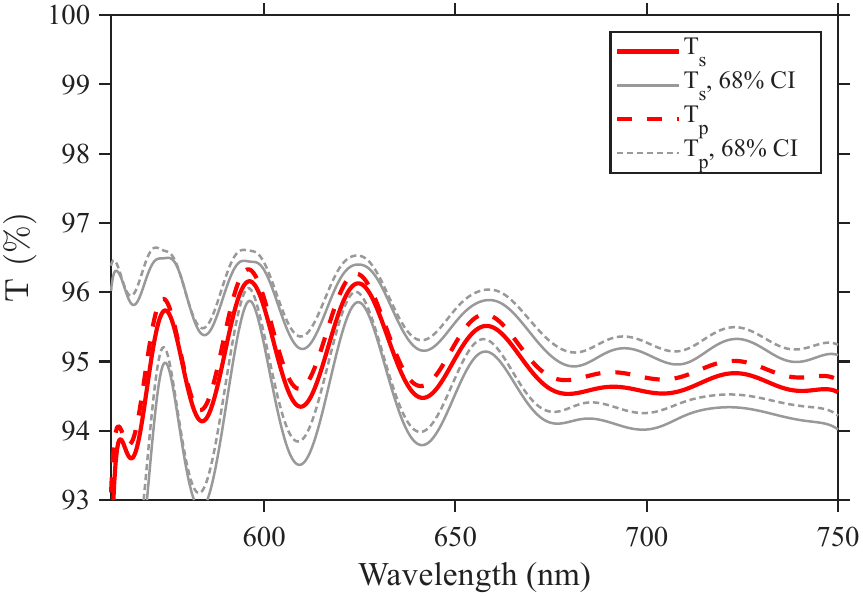}
    \subcaption{}
    \label{fig:final_montecarlo_t}
  \end{subfigure}

  \caption{
Monte Carlo tolerance analysis of (a) reflectance and (b) transmittance. Solid curves show the expected performance, while gray curves indicate a 68\% confidence interval.
}
\label{fig:final_montecarlo}
\end{figure}
The corresponding Monte Carlo reflectance and transmittance results are shown in Fig.~\ref{fig:final_montecarlo}. The expected spectral performance remains close to the nominal design, and the confidence intervals indicate only modest variation in both bands. Together, these results demonstrate that the final phase-optimized design maintains both high spectral performance and low chromatic wavefront-error sensitivity under realistic fabrication tolerances.

\label{sec:montecarlo}

\section{Implications for coronagraph performance}
\label{sec:contrast5}

To evaluate the impact of chromatic coating-induced WFE on coronagraph performance, we interpret the wavelength-dependent wavefront error variability \(\sigma_{\mathrm{WFE}}(\lambda)\), derived from the Monte Carlo phase sensitivity analysis described in Sec.~\ref{sec:phaseopt}, as a proxy for the spatial WFE across the dichroic. The resulting chromatic phase aberrations were propagated through a simulation of a charge-6 vector vortex coronagraph (VVC) implemented using HCIPy. \cite{roy2026lazuli,Por2018HCIPy}

\begin{figure}[ht]
  \centering
  \begin{subfigure}[t]{.9\textwidth}
    \centering
    \includegraphics[width=\textwidth]{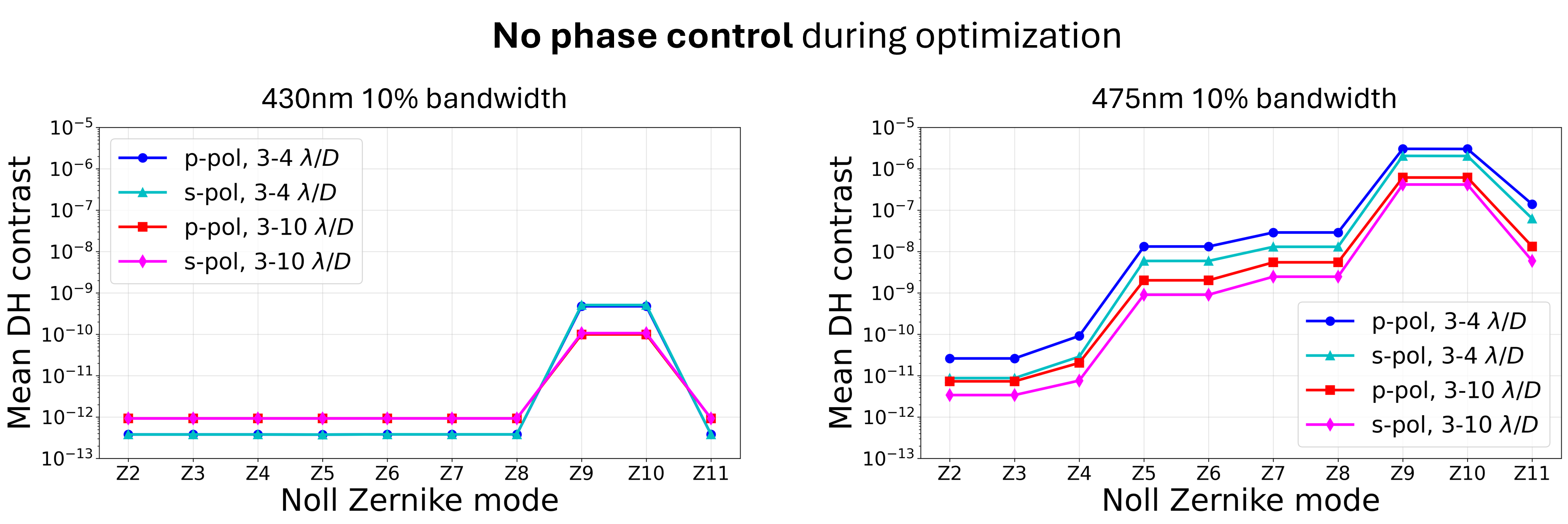}
    \subcaption{}
    \label{fig:nophaseopt_zvc}
  \end{subfigure}\hfill
  
  \begin{subfigure}[t]{.9\textwidth}
    \centering
    \includegraphics[width=\textwidth]{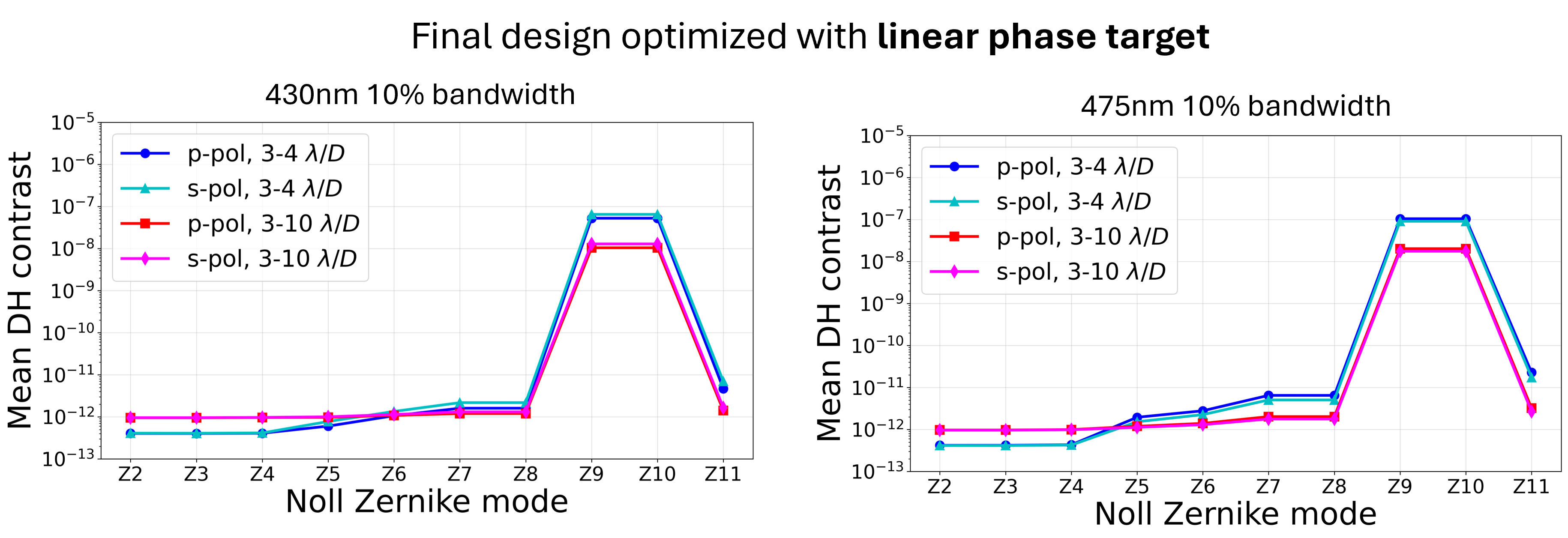}
    \subcaption{}
    \label{fig:mydesign_zvc}
  \end{subfigure}
\caption{Coronagraph sensitivity of the charge-6 VVC to coating-induced Zernike phase aberrations for (a) the baseline design and (b) the final linear phase optimized design. Mean dark-hole contrast is shown for both the \(3\)--\(4~\lambda/D\) and \(3\)--\(10~\lambda/D\) regions using a 10\% spectral bandwidth at central wavelengths of 430~nm and 475~nm. Contrast differences between s- and p-polarization increase in the baseline design's region of rapid spectral phase variation.}
\label{fig:zernike_contrast}
\end{figure}

Two representative wavelength bands centered at 430~nm and 475~nm were selected for analysis. As shown in Fig.~\ref{fig:sig_compare}, the design optimized without a phase constraint exhibits relatively little variation in \(\sigma_{\mathrm{WFE}}(\lambda)\) across the 430~nm band, whereas a pronounced spectral feature is present near 475~nm. For each band, individual Zernike modes were scaled by \(\sigma_{\mathrm{WFE}}(\lambda)\), and the mean dark-hole (DH) contrast was evaluated over a 10\% spectral bandwidth using both a \(3\)--\(4~\lambda/D\) and \(3\)--\(10~\lambda/D\) dark-hole region. Figure~\ref{fig:zernike_contrast} compares the resulting coronagraph sensitivity for the baseline dichroic design and the final phase-optimized design. Results are shown separately for s- and p-polarized light to illustrate the influence of polarization-dependent coating phase variations on coronagraph performance. Although the overall contrast trends are similar for both polarization states, small differences occur due to the polarization dependence of the coating-induced wavefront error shown in Fig.~\ref{fig:final_wfe}.

\begin{figure}[ht]
    \centering
    \includegraphics[width=\textwidth]{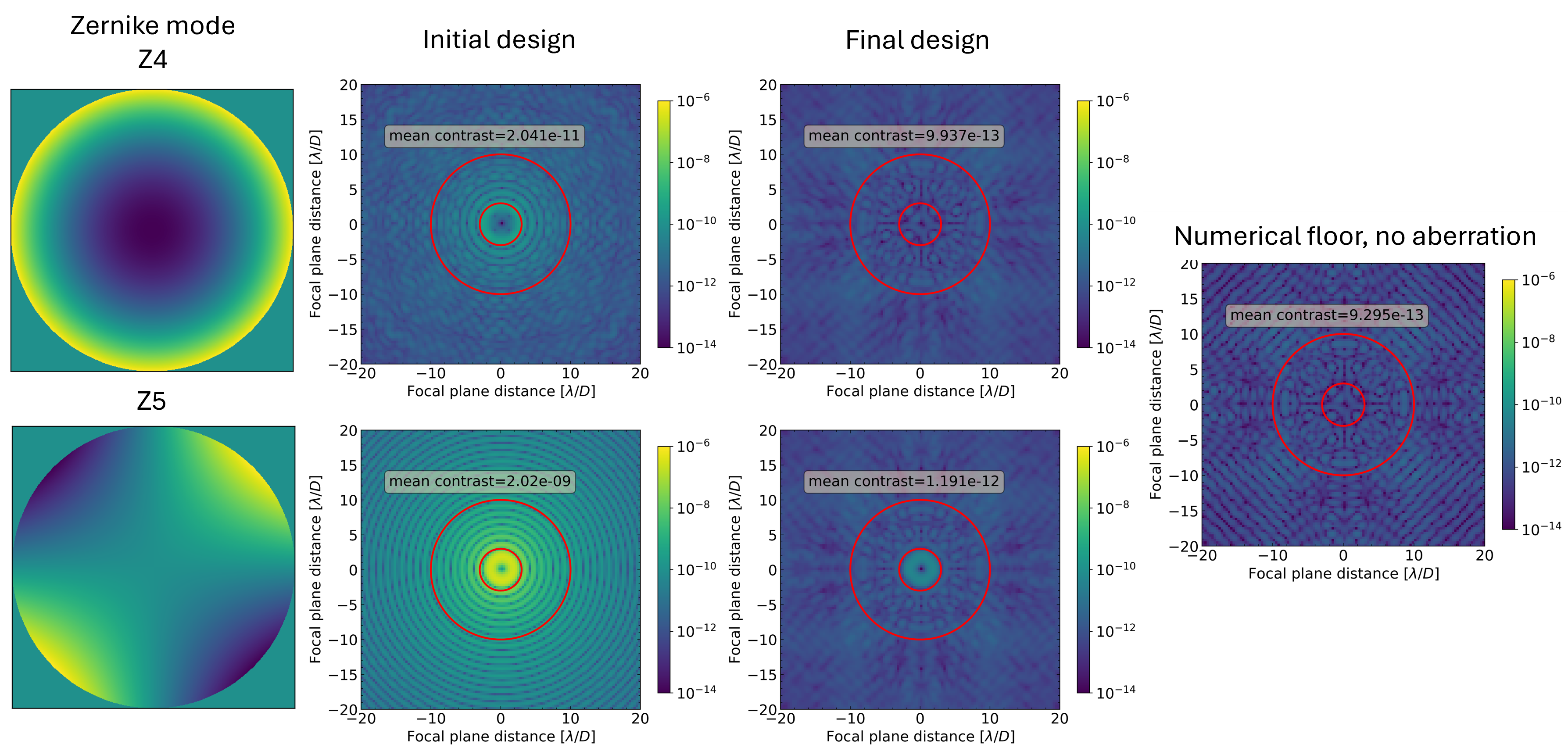}
    \caption{Representative 475~nm dark-hole intensity distributions for Noll Z4 and Z5, evaluated with a charge-6 VVC over a 10\% spectral bandwidth. Red circles mark the \(3\)--\(10~\lambda/D\) region used to calculate the reported mean contrast values; the far-right panel shows the numerical floor with no applied aberration.}
    \label{fig:zernike_psf}
\end{figure}

The results indicate that the design optimized for linear reflected phase achieves ${<}10^{-10}$ broadband contrast if the chromatic phase aberrations manifest as low-order aberrations. At 475~nm, the baseline design shows a steady degradation in dark-hole contrast from Noll Zernike modes Z5 through Z11, reaching contrast levels above \(10^{-6}\) for the most sensitive modes. In contrast, the phase-optimized design suppresses the response of most low-order modes to the \(10^{-12}\)--\(10^{-11}\) level, with significant sensitivity remaining only for the trefoil modes, Noll Z9 and Z10. This behavior is consistent with the charge dependence of vortex coronagraph leakage described by Ruane et al., which shows that rejection of a Zernike mode depends on the vortex charge relative to the mode's radial and azimuthal orders. \cite{Ruane_2018} Furthermore, the phase-optimized design exhibits nearly identical contrast performance at 430~nm and 475~nm, whereas the baseline design displays a strong wavelength dependence. This suggests that correction of these aberrations using a deformable mirror would be feasible for the phase-optimized design. To further illustrate this contrast improvement, Fig.~\ref{fig:zernike_psf} compares representative coronagraphic dark-hole intensity distributions at 475~nm for defocus (Z4) and horizontal astigmatism (Z5). These modes were selected because dichroic-induced wavefront error is expected to couple most strongly into low-order aberrations.

\newpage

\begin{figure}[ht]
  \centering
  \begin{subfigure}[t]{.5\textwidth}
    \centering
    \includegraphics[width=\textwidth]{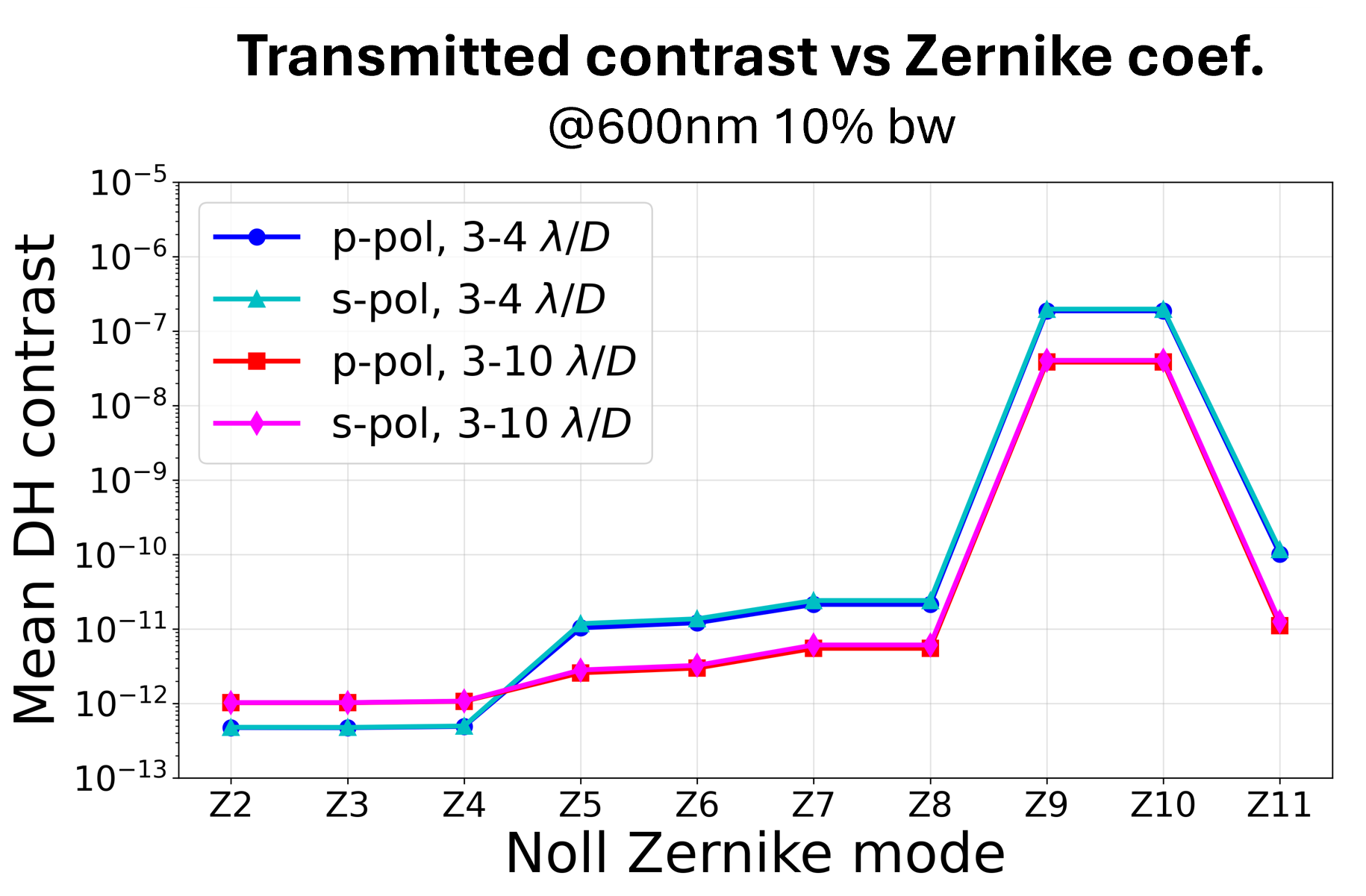}
    \subcaption{}
    \label{fig:trans_600}
  \end{subfigure}\hfill
  \begin{subfigure}[t]{.5\textwidth}
    \centering
    \includegraphics[width=\textwidth]{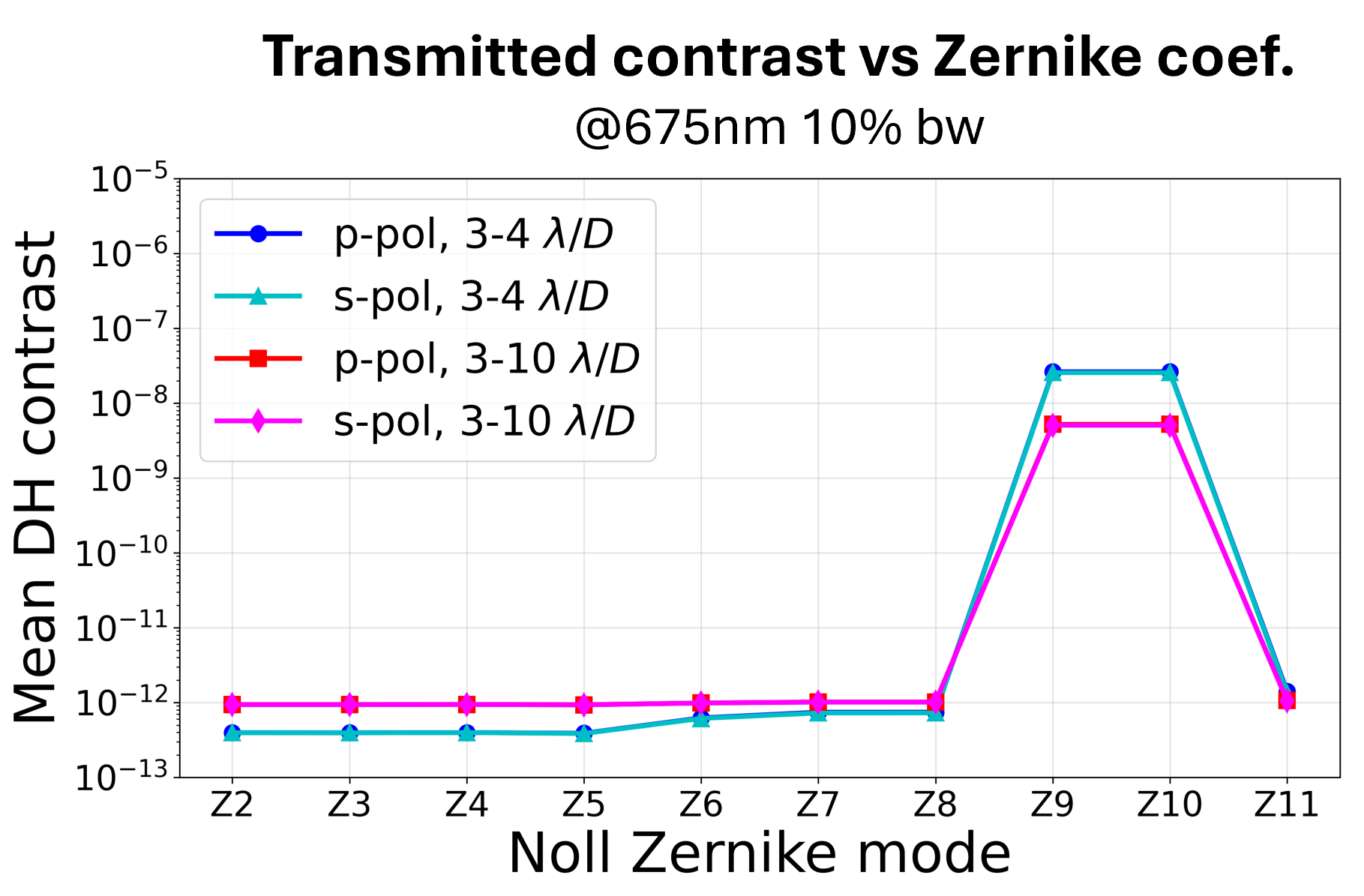}
    \subcaption{}
    \label{fig:trans_675}
  \end{subfigure}
\caption{Transmitted-channel coronagraph sensitivity at (a) 600~nm and (b) 675~nm. Mean dark-hole contrast is shown for individual Zernike modes using both s- and p-polarization wavefront-error sensitivities. Results are reported for the $3$--$4~\lambda/D$ and $3$--$10~\lambda/D$ dark-hole regions.}
\label{fig:zernike_trans}
\end{figure}

While the preceding analysis focused on the reflected channel, the transmitted channel was also evaluated. Fig.~\ref{fig:zernike_trans} compares coronagraph sensitivity for representative wavelengths within the transmission band. The transmitted band exhibits similar trends in Zernike mode sensitivity when compared to the reflected band. Near the transition region at 600~nm, several low-order aberrations (Z5--Z8) exhibit elevated sensitivity to chromatic wavefront error. At 675~nm, farther from the spectral transition, the response is dominated primarily by the trefoil terms (Z9 and Z10). In both cases, the contrast difference between s- and p-polarization remains small, indicating that polarization-dependent effects are well controlled throughout the transmission band. This is consistent with the results shown in Fig.~\ref{fig:final_wfe_d}, where the orthogonal polarization states have nearly equal WFE sensitivity across the transmitted band.

\newpage
\section{Conclusion}

A low-AOI dichroic mirror was designed and analyzed for use in a high-contrast imaging system, with particular emphasis placed on polarization-dependent phase behavior and its impact on high-contrast imaging performance. The final coating design successfully met the required reflection and transmission band specifications, while maintaining low diattenuation and reduced polarization-dependent phase variations across the operating wavelength range. 

In addition to conventional spectral performance metrics, the reflected coating phase response was incorporated directly into the optimization process. Monte Carlo sensitivity analysis demonstrated that designs optimized for a more linear spectral reflected phase response exhibit substantially lower wavefront error and reduced sensitivity to manufacturing-induced thickness variations. The final phase-optimized coating produced a smoother spectral wavefront-error response and eliminated localized features associated with increased chromatic sensitivity.

To evaluate the system-level implications of these improvements, we propagated the coating-induced chromatic wavefront error through a model of the vector vortex coronagraph. These results showed that linear phase optimization not only reduces coating-induced wavefront error, but also improves coronagraph sensitivity to chromatic aberrations by reducing chromatic residuals within the dark hole. These findings demonstrate the importance of considering the spectral variation of reflected phase during the design of dichroic coatings intended for high-contrast imaging applications. Coronagraph simulations performed for s- and p-polarization demonstrated that these residual polarization effects produce only modest differences in dark-hole contrast, confirming that the low-AOI design effectively minimizes polarization sensitivity.


Fabrication of the designed dichroic is in progress. Once fabrication is complete, the spectral performance, Mueller matrix\cite{Ashcraft2024Gromit}, and WFE will be characterized experimentally.\cite{euclid_measure} This prototype will be used on the Space Coronagraph Optical Bench (SCoOB) to determine contrast performance.\cite{Anche2024SCoOBPolAberrations,vangorkom2022spacecoronagraphopticalbench,nikoposter} Comparison of modeled and experimental results will assess the accuracy of the optimization and tolerancing procedure discussed in this paper, as well as indicate potential ways to improve analysis techniques.

\acknowledgments
Portions of this research were supported by funding from the Technology Research Initiative Fund (TRIF) of the Arizona Board of Regents and by generous anonymous philanthropic donations to the Steward Observatory of the College of Science at the University of Arizona. This research received support through Schmidt Sciences.

\bibliography{mybib}
\bibliographystyle{spiebib}

\end{document}